\pdfoutput=1
\documentclass[conference]{IEEEtran}
\usepackage{amsmath,amssymb,amsfonts}
\usepackage{mathtools}
\usepackage{dsfont}
\usepackage{soul}
\usepackage[dvipsnames]{xcolor}

\usepackage[export]{adjustbox}
\usepackage{caption}
\usepackage{algorithmic}
\usepackage{lettrine}
\usepackage{graphicx}
\usepackage{textcomp}
\usepackage{cancel}
\usepackage{subcaption}
\usepackage{float}
\usepackage{hyperref}
\usepackage{titlesec}
\usepackage[english]{babel}
\usepackage{amsthm}
\usepackage[ruled,vlined]{algorithm2e}
\usepackage[shortlabels]{enumitem}

\theoremstyle{definition}

\DeclareGraphicsExtensions{.png,.pdf,.jpg}
\usepackage{subcaption}
\usepackage{array}
\usepackage{multicol}
\usepackage{cite}
\usepackage{xcolor}
\usepackage{array,multirow,textcomp}
\theoremstyle{definition}

\usepackage{subcaption}
\usepackage{geometry} 
\usepackage{authblk}
\usepackage{etoolbox}
\usepackage{graphicx}%
\begin{document}
\title{RAG-Enabled Intent Reasoning for Application-Network Interaction \\
\thanks{Mohammed S. Elbamby contributed to this work as a member of Nokia Bell Labs Finland until February 2024.}}
\author[$1$]{Salwa Mostafa}
\author[$2$]{Mohamed K. Abdel-Aziz}
\author[$3$]{Mohammed S. Elbamby}
\author[$1$]{Mehdi~Bennis}
\affil[$1$]{Centre for Wireless Communications, University of Oulu, FI-90014 Oulu, Finland.}
\affil[$2$]{Nokia Bell Labs,  Espoo, Finland}
\affil[$3$]{Telefonica Research, Barcelona, Spain}
\affil[$1$]{salwa.mostafa, mehdi.bennis@oulu.fi}
\affil[$2$]{mohamed.3.abdelaziz@nokia-bell-labs.com}
\affil[$3$]{mohammed.elbamby@telefonica.com}
\maketitle

\begin{abstract}
Intent-based network (IBN) is a promising solution to automate network operation and management. IBN aims to offer human-tailored network interaction, allowing the network to communicate in a way that aligns with the network users' language, rather than requiring the network users to understand the technical language of the network/devices. Nowadays, different applications interact with the network, each with its own specialized needs and domain language. Creating semantic languages (i.e., ontology-based languages) and associating them with each application to facilitate intent translation lacks technical expertise and is neither practical nor scalable. To tackle the aforementioned problem, we propose a context-aware AI framework that utilizes machine reasoning (MR), retrieval augmented generation (RAG), and generative AI technologies to interpret intents from different applications and generate structured network intents. The proposed framework allows for generalized/domain-specific intent expression and overcomes the drawbacks of large language models (LLMs) and vanilla-RAG framework. The experimental results show that our proposed intent-RAG framework outperforms the LLM and vanilla-RAG framework in intent translation.
\end{abstract}


\section{Introduction}

The plethora of evolving digital applications and services such as immersive multimedia and connected vehicles drives network operators toward automating network operations and management. Intent-based network (IBN) has recently been introduced as a promising solution to automate networks~\cite {leivadeas2022survey}. IBN relies on receiving application/service demand in an abstract and high-level description language without specifying how to achieve the desired outcome, which is called {\em intent}. IBN is responsible for translating the received intent into low-level configuration language to be executed on network functions and devices. This gives high freedom to network operators and service providers to manage and optimize their network resources. However, translating the high-level expressed intents to low-level network configuration language is a challenging problem since various applications interact with the network, each with its specialized needs and domain language. Traditional translation methods rely on creating an ontology-based language for each application to facilitate intent translation, which lacks technical expertise and is neither practical nor scalable. To tackle this problem, various efforts from both industry and academia have been made to introduce artificial intelligence (AI) and machine reasoning (MR) to mobile networks. AI and MR have powerful capabilities to interpret different intent languages and deploy translated intents into network configurations through compilers. 

The studies in~\cite{hui2020intent,jacobs2018refining} focused on machine learning translators, where machine learning techniques such as deep neural networks (DNNs) are used to provide keyword classification. The study in~\cite{hui2020intent} proposed an intent-defined optical network (IDON) framework that uses a deep neural network to extract classes, providing the relationship between service characteristics and intentions. Moreover, word embedding and convolutional neural network (CNN) approaches are adopted to get similarity scores between keywords. The work in~\cite{jacobs2018refining} proposed a scheme based on recurrent neural networks (RNN) for a sequence-to-sequence learning model. The scheme predicts what kind of keyword to expect and extracts semantic correlations between different items of the intent. The works in~\cite{toy2021intent,alsudais2017hey} proposed keywords-based translators, which specify specific services based on certain keywords used in the intent and their associated correlation keywords. Unfortunately, the above-mentioned translators rely only on keyword detection and cannot reason over different domain languages to interpret the intent from a specific domain perspective and utilize the network knowledge data to translate intents to get the right configuration for different applications. Moreover, these approaches are not suitable for intents expressed by non-technical users, as their expressed intent can be application-specific and challenging to map to a specific service through keyword mapping.

Recently, LLMs based on transformers such as chatGPT, Gemini, etc. have shown great potential in improving IBN performance~\cite{fuad2024intent,dzeparoska2023llm,mekrache2024intent,lin2023appleseed} due to their powerful capabilities in performing natural language processing (NLP) tasks such as understanding, generating, and classifying text data. The studies in~\cite{fuad2024intent,dzeparoska2023llm,mekrache2024intent,lin2023appleseed} adapted LLMs and MR in intent translation frameworks. These frameworks rely on classifying intents and extracting actions based on keywords in the expressed intents. The authors assume that the user intents are provided by technical experts and that the user's intents contain the required technical information to configure the network, which does not generalize these frameworks for non-technical user intents. Besides, it may induce inaccurate and outdated information in network configuration as LLMs rely on their trained data. 

To solve the application-network interaction problem, in this paper, we propose a framework that utilizes application domain context, RAG, LLM model with reasoning capabilities, and few-shot learning to reason over high-level application language intents (i.e., technical and non-technical) and translate them to low-level network configuration language. RAG technique is considered a class of LLM applications that use external data to augment the LLM’s context. It helps augment knowledge-intensive tasks and ensures LLMs have up-to-date knowledge. LLM models with reasoning capabilities can provide accurate translation and reasoning over the domain language of different applications. Few-shot learning~\cite{agrawal2024mindful} is a framework that trains the LLM through a very small number of examples via a prompt. It overcomes the hallucination issue and improves the LLMs' responses without retraining or fine-tuning, as retraining or fine-tuning the LLM is costly and consumes a lot of resources. Our main contribution can be summarized as follows:

\begin{itemize}

\item We investigate an application-network interaction problem, where different applications aim to interact in their domain language with the network. The network must understand and translate their requested services/demands into network configuration language.  

\item We propose a context-aware AI framework that acts as a functionality in the vertical application layer able to interpret different languages from various applications and translate them to a common language understood by the network service provider. The framework relies on the application domain context, RAG model, reasoning LLM model, and few-shot learning to preserve data privacy, provide an accurate translation, and reduce hallucination in the translated intents. 

\item Experimental results show the performance improvement of our proposed Intent-RAG framework compared to LLM and vanilla-RAG benchmarks.

\end{itemize}

The rest of the paper is organized as follows: In Section~\ref{Problem}, we state the application network interaction problem and the proposed AI solution framework. Section~\ref{simulation} provides the experimental setup and results. Finally, we conclude the paper in Section~\ref{conclusion}.

%\vspace{10mm}

\section{Problem Statement and Proposed Framework}~\label{Problem}

\subsection{Problem Statement}~\label{ProblemStatement}

We consider a scenario, where different applications, such as virtual reality (VR)~\cite{TR126999}, augmented reality (AR)~\cite{TR126998}, vehicular to everything (V2X)~\cite{TR126985}, etc. aim to consume network services with certain expectations/requirements expressed in the domain language of these applications. Since application users mostly are not network experts, they look forward to expressing their expectations from the network service provider in their domain language. By {\em{'domain language'}} here, we mean that each application service request is expressed to the network in terminologies of the application expectation or operation language. For example, for the VR application, a user can express his intent as \textit{"I want VR without motion sickness."}. The operation network needs to interpret their demands/expectations and translate them into a network language (i.e., quality of service (QoS), quality of experience (QoE), and configurations) to deploy a network policy to activate network devices and functions. Thus, this study investigates the interaction problem between different applications and network service providers.

\subsection{Problem Formulation}

We consider the problem of translating high-level, application-specific intents into structured network intents that can be directly interpreted by the network. Let 
\(\mathcal{I}\) denote the \emph{intent space} containing natural language intents expressed by users or applications, and let \(\mathcal{N}\) denote the \emph{structured network intent space} defined according to standardized information models (e.g., 3GPP, TMF). The network knowledge is stored in an external knowledge base \(\mathcal{K}\), which contains domain-specific technical information extracted from standards, configuration manuals, and service catalogs. The translation task can be formulated as:
\begin{equation}
    \hat{n} = \arg\max_{n \in \mathcal{N}} P(n \mid i, \mathcal{K}),
\end{equation}
where \( P(n \mid i, \mathcal{K}) \) denotes the conditional probability of generating a structured network intent \( n \) given the natural language intent \( i \) and the knowledge base \(\mathcal{K}\).

Given an input intent \( i \in \mathcal{I} \), the goal of the proposed framework is to infer a structured network intent \( n \in \mathcal{N} \) conditioned on both the intent \( i \) and the relevant knowledge \( \mathcal{K} \).

The proposed framework consists of two major stages:
\begin{itemize}
    \item \textbf{Intent Refinement:} A mapping 
    \(\phi: \mathcal{I} \to \tilde{\mathcal{I}}\),
    where \(\tilde{\mathcal{I}}\) is a well-defined intent space obtained through ambiguity resolution and context enrichment.
    \item \textbf{Structured Intent Creator:} A mapping 
    \( f_\theta: (\tilde{\mathcal{I}}, \mathcal{K}) \to \mathcal{N} \),
    parameterized by \(\theta\) (e.g., the language model and prompting strategy), which generates the structured intent.
\end{itemize}

The optimal structured intent is obtained by minimizing a translation loss function \( \mathcal{L} \) between the generated intent and the ground-truth structured intent \( n^\ast \):
\begin{equation}
    \theta^\ast = \arg\min_{\theta} 
    \mathbb{E}_{(i, n^\ast) \sim \mathcal{D}} 
    \big[\mathcal{L}(f_\theta(\phi(i), \mathcal{K}), n^\ast)\big],
\end{equation}
where \( \mathcal{D} \) is the training or evaluation dataset, and 
\(\mathcal{L}\) may include semantic similarity, schema conformity, or faithfulness scores.

\subsection{Proposed Framework}~\label{ProposedFramework}

\begin{figure*}
\centering
\includegraphics[width=\textwidth, height = 4.8 in]{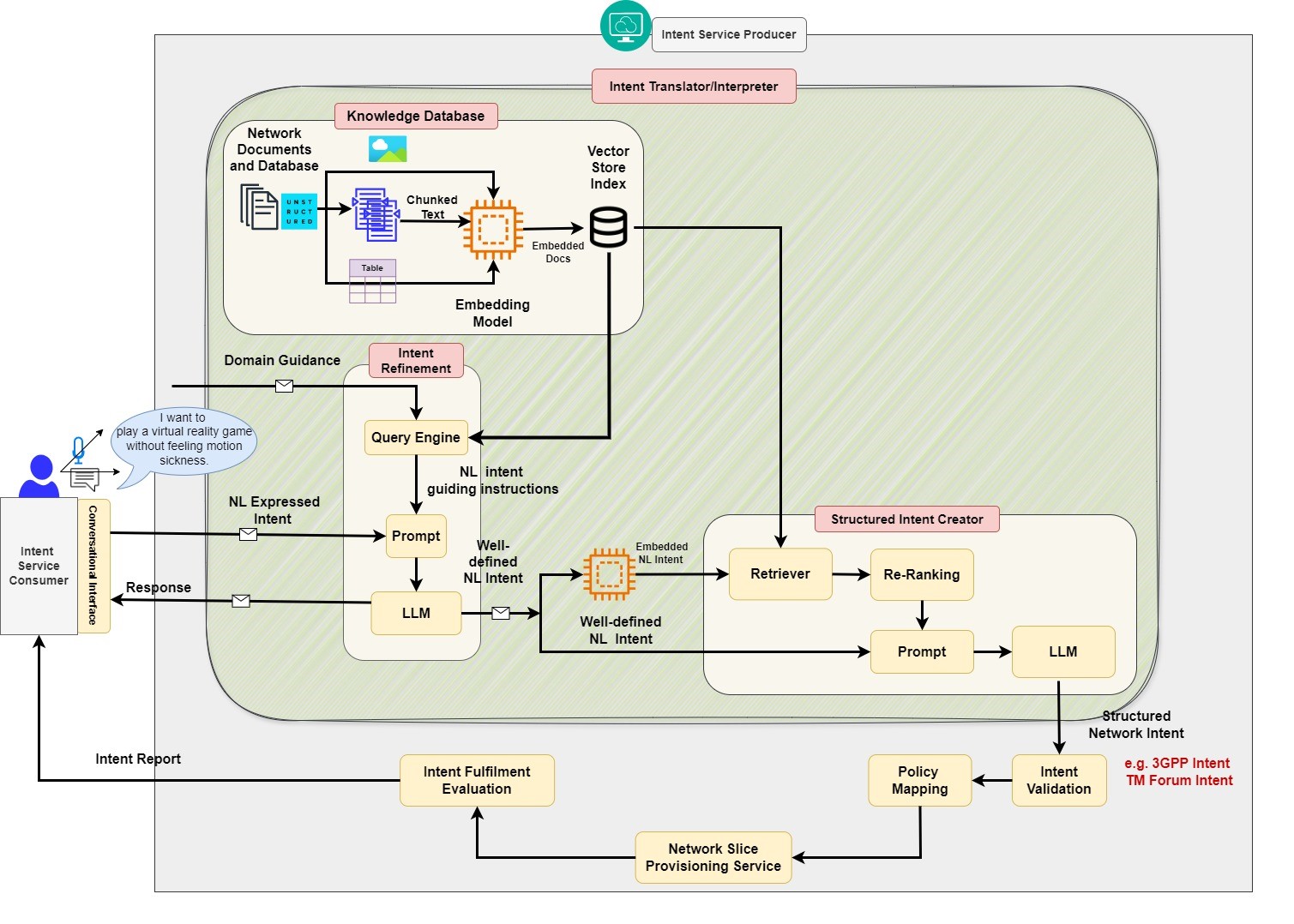}
\caption{Intent-RAG Framework for Intent Interpreting.}
\label{framework}
\end{figure*}

To tackle the application-network interaction problem stated above, we propose the context-aware AI framework shown in Fig.~\ref{framework} called {\em Intent-RAG}. The framework utilizes application domain context, RAG, LLM models with reasoning capabilities, and few-shot learning to build a reasoning framework over high-level application language intents. A conversational interface is considered to provide bidirectional communication between the user application and the network. It receives the user intent as input and sends the response back to the user. After receiving the user intent, it converts it to a text format if sent in another format (i.e., voice, image, etc.). The framework works as follows: First, a knowledge database is created by collecting technical documents necessary for the intent translation of the target application. These documents are then chunked, embedded, and stored in a vector database. Second, the user application sends its high-level generic intent to an intent refinement functional block, which converts it to a well-defined intent (i.e., intent contains essential information to search the knowledge database). Third, a structured intent creator extracts the relevant information to a well-defined intent from the knowledge database, ranks them based on high similarity, and generates a structured network intent. 

\begin{multicols}{2}
\end{multicols}
\begin{multicols}{2}
\end{multicols}

The framework consists of the following functional entities:

\begin{itemize}

\item {\textbf{Knowledge Database:}} It prepares the network knowledge domain to be queried and retrieved. Since the technical documents have special formats, optimizing the processing of these documents and storing them in a database is very critical to the performance of the AI framework; thus, here we show the details of this functional block:

\begin{enumerate}

\item {\em{Documents Loader:}} We implement the unstructured documentation~\cite{unstructured} to load the technical documents of the target application domain and identify the different modalities in their context, such as text, table, image, etc.. For better retrieval performance, we also implement a summarizing model for each modality, where after separating each module we create a summary of it.

\item {\em{Text Splitter:}} It chunks the text modalities into small nodes/pieces to be queried and retrieved. 

\item {\em{Embedding Model:}} It generates a numerical representation of each modality to perform a similarity/semantic search. It also embeds the well-defined intent before the search operation. 

\item {\em{Vector Store Index:}} It stores the vector embedding representations of the technical documents and databases so that different types of searches, such as semantic, similarity, etc. can be applied immediately to retrieve relevant technical information to the received intent.

\end{enumerate}

\item {\textbf{Intent Refinement:}} It is responsible for refining the generic user application intent to a more clearly well-defined intent for a specific service domain. It reduces ambiguity in the generic expressed intents through guidance from the service domain. It acts as an intent classifier by analyzing the application user’s intent to specify the application type and need. It consists of three functional blocks: 

\begin{enumerate}

\item {\em{Query Engine:}} It is responsible for querying the network knowledge database (i.e., stored in vector store index) and providing guidance instructions to clarify ambiguous intents based on the service domain instruction.

\item {\em{Prompt:}} The prompt takes two inputs: the user application's natural language (NL) intent and the NL intent guiding instructions. The user application's NL intent is expressed in a natural language in the domain of the service consumer (i.e., can be technical or non-technical). The guided instruction is given by the service provider domain based on the scope of the requested services to handle ambiguous intents.

\item {\em{LLM:}} The LLM has reasoning capabilities and takes the prompt as input and produces a well-defined intent that can be used to query the network knowledge database to extract the proper configuration and execution plans for application services. A well-defined intent is a statement containing essential technical information to query the network knowledge database. 

\end{enumerate}

\item {\textbf{Structured Intent Creator:}} It retrieves information from the network knowledge database based on the well-defined NL intent and creates a structured network intent:

\begin{enumerate}

\item {\em{Retriever:}} It retrieves relevant context from the vector store index when a well-defined intent is given as input. The retriever's performance highly depends on the chunking method used to chunk the technical document and the embedding model.

\item {\em{Re-Ranking:}} It reorders the retrieved context information from the retrieval based on relevancy. The top-ranked contexts are then input to the LLM through the prompt.

\item {\em{Prompt:}} It takes the well-defined intent and the top-ranked retrieved relevant contexts as input and then guides the LLM to output a structured network intent defined by the standards developing organizations (SDOs) (i.e., 3GPP, TMF, etc..).

\item {\em{LLM:}} It takes the prompt as input and produces a structured network intent following the information models defined by SDOs (i.e., 3GPP, TMF, etc.).

\end{enumerate}

\end{itemize}

The structured network intent is then passed to an intent validation to monitor whether there are conflicts in the translated intent or not. Afterward, the policy mapping function maps the network intent to network policy. It can be predefined blueprints/templates that are predefined and map specific ordered virtual network functions (VNFs) to form a Service Function Chain (SFC). These SFC blueprints can be stored in a database that corresponds to a combination of QoS levels. Then, the policy is passed to the provisioning service. Afterward, the intent fulfillment evaluator sends an evaluation report to the intent consumer about the fulfillment status of the intent~\footnote{The details of this part are out of the scope of the current work.}.

\section{Intent-RAG: Implementation AND Evaluation}~\label{simulation}

This section illustrates the detailed implementation of our proposed intent-RAG framework. We also validate and evaluate its efficacy for an IBN scenario and compare it to using an LLM model directly and a vanilla-RAG framework as a benchmark. Both intent-RAG and vanilla-RAG frameworks are implemented using {\textbf{LlamaIndex}} and {\textbf{RAGAS}} evaluators.  In the IBN scenario, we consider the service and traffic scenarios defined in the following network document~\cite{TR103702}, where the performance requirements and key factors are described for different service and traffic scenarios. The network document is fed to the network knowledge database to form structured network intents based on the information in the document. Note that the implementation parameters are chosen based on experimental trials. The detailed implementation is as follows:

\subsection{Implementation}
\subsubsection{\bf{Intent-RAG}}
\begin{itemize}
\item {\textbf{The knowledge database is constructed as follows:}}
\begin{enumerate}
\item The unstructured tool~\cite{unstructured} is used to preprocess the unstructured documents of the network. It extracts the different modalities, such as texts, tables, images,..,etc..
\item The extracted text modal is then chunked by a SentenceSplitter with chunk size $128$ and chunk overlap $10$.
\item A summarization pipeline which consists of a prompt and pre-trained summarization LLM is used to produce a summary for each extracted modal. \\
-	The prompt for the text modal is designed as {\bf{“You are an assistant tasked with summarizing text. Give a concise summary of the text {\em\{text chunk\}} and a title to it. Please provide the summary in a string format.”}} \\
-	The prompt for the table modal is designed as {\bf{“You are an assistant tasked with summarizing tables. Give a very concise summary about what is this table {\em\{table\}} about? Please provide the summary in a string format.”}}

\item The original modalities and their summarization are embedded using the OpenAI embedding model “text-embedding-ada-002” and stored in~\href{https://www.trychroma.com/}{Chroma} database.
\end{enumerate}
\item {\textbf{The intent refinement is constructed as follows:}} 
\begin{enumerate}
\item We consider an IBN use case, where the service/traffic consumer is an application user requesting different service/traffic scenarios in a generic form from an application service provider. Thus, the service domain gives the following domain instructions to the query engine “Please list all the service/traffic scenarios that can be provided to our customers”. 
\item  The query engine accesses the knowledge database and gives a list of all supported applications “The list of supported service/traffic scenarios are Urban macro, Airplanes connectivity, 4K On-Demand Video, etc..” and passes it to the prompt.
\item  The prompt is designed as {\bf{"You are an expert network service provider who can predict the service/traffic scenario demanded by network users. Predict the most relevant service/traffic scenario requesting the following service demand {\em\{NL expressed intent\}} from the following list {\em\{query engine output\}}. Some examples are given below: intent: 4K On Demand Video. Service/traffic scenario: 4K On Demand Video. intent: I want internet access with fast browsing service in the airoplane. Service/traffic scenario: Airplanes connectivity. Write a clear and conscious output giving only the chosen service/traffic scenario. Do not put any introductory phrases, commentary, or explanations." }}
\item The prompt is passed to an OpenAI LLM model with reasoning capabilities such as “o1-mini”. Assume that the NL expressed intent passed to the LLM is “I want to play a virtual reality game without feeling motion sickness”. The LLM response is “Scenario Type : 3K Cloud VR (Game)”. 
\end{enumerate}

\item {\textbf{The structured intent creator is constructed as follows:}}
\begin{enumerate}
\item The retriever is set to retrieve the top 6 information nodes related to the well-defined NL intent.
\item The reranker reorders them and returns the top 3 relevant contexts based on the Coherent Reranker scheme, which gives better search results than keyword-based, semantic-based, and embedding-based semantic search.
\item The prompt is designed as {\bf{"Provided the following context information {\em\{reranker output context\}}. Given only this information, Please provide the performance recommendations metrics and their values to the scenario {\em\{well-defined NL intent\}} in the following format {3GPP/TMF format given}. Some examples are given below: Scenario Type: 4K On Demand Video, Key Performance Factors: Data Rate/Throughput (downlink):30 Mbps, Delay: RTT $< 100$ ms, Packet Loss Rate:$10^{-3}$, Resolution: 4K, Coverage Level CSI RSRP:$-113$ dBm, Coverage Quality CSI SINR: $-2$ dB.”}}.
\item The instruction OpenAI LLM “gpt-3.5-turbo-instruct” is used to produce the output according to the format given.

\end{enumerate}
\end{itemize}

\subsubsection{\bf{Vanilla-RAG}}

The knowledge database is constructed by loading the technical documents with a simple directory reader from LlamaIndex. Then, the documents are chunked using a SentenceWindowNodeParser with window size~$3$. Afterward, the chunks are embedded by the embedding model “text-embedding-ada-002” and stored in the Chroma database. A retriever is implemented to retrieve the top 3 related contexts to the intent. Finally, the retrieved context and intent are passed through a prompt to an instructed LLM to generate the structured network intent. Note that the prompt is designed similarly to the prompt for the intent RAG for a fair comparison.
 
\subsection{Evaluation}

The experiment is conducted on an intent interpretation use case, where a set of network user applications express their intents in their own domain's language to the network.  Each user application generates an expectation/demand for the network in a human language. The network has to interpret and translate the application intents to structured network intents through the proposed framework. To evaluate the performance of intent-RAG and vanilla-RAG frameworks for our use case, we use both human evaluation and the RAGAS evaluator~\cite{es2023ragas}. The dataset used for evaluation includes both technical and non-technical user intents and is available in \cite{dataset}. The human evaluation is presented in Table~\ref{table_2}, where the ground truth is extracted manually from the technical documents. 

For the RAGAS evaluator, the evaluation metrics context precision, context recall, and context entity recall reflect the performance of the retrieval part in the framework, where context precision measures how relevant the retrieved context is to the user intent, whereas context recall measures the retriever’s ability to retrieve all necessary information required to construct the network intent. Context entity recall measures the recall of the retrieved context, based on the number of entities present in both ground truths and contexts relative to the number of entities present in the ground truths alone. The evaluation metrics answer relevancy, answer correctness, and faithfulness reflect the performance of the generative part in the framework. The answer relevancy measures how relevant the framework responds to the user intent. The answer correctness measures the accuracy of the generated response when compared to the ground truth. The evaluation metric faithfulness measures hallucinations in the framework.

\begin{multicols}{2}
\end{multicols}
\begin{table*}[h]
\caption{Comparison between LLM (no-RAG), vanilla-RAG, and intent-RAG based on human evaluation}
\centering
\begin{tabular}{|>{\centering\arraybackslash}p{1.5cm}|>{\centering\arraybackslash}p{7.5cm}|>{\centering\arraybackslash}p{7.5cm}|}
\hline
User Intent &  4K On Demand Video  & I want to play a virtual reality game without feeling motion sickness    \\ 
\hline
Vanilla-RAG Output &  Scenario Type: 4K On Demand Video
Key Performance Factors: - Data Rate/Throughput (downlink): $30$ Mbps
- Delay: RTT $< 50$ ms - Frame Rate: $60$ fps - Resolution: $4K (3 840 \times 2 160)$
- Operation Response Delay: $50$ ms - Coverage Quality CSI SINR: $-2$ dB & Scenario Type: Cloud VR Interactive VR Games, Key Performance Factors/Performance Recommendations: Bandwidth: $100$ Mbps (downlink), Packet Loss Rate: $< 0.01\%$ (UDP), Network Delay: $< 25$ ms, MTP Delay (ms): $< 80$ ms (good experience), Coverage level CSI RSRP: $-107$ dBm, Coverage level CSI SINR: 2 dB.\\
\hline
Intent-RAG Output  & Scenario Type: 4K On Demand Video, Key Performance Factors: Data Rate/Throughput (downlink): $30$ Mbps, Delay: RTT $< 100$ ms, Packet Loss Rate:$10^{-3}$, Resolution: $4$K, Coverage Level CSI RSRP: $-113$ dBm, Coverage Quality CSI SINR: $-2$ dB. & Scenario Type: 3K Cloud VR (Game), Key Performance Factors: Data Rate/Throughput (downlink): $100$ Mbps, Delay: RTT $< 25$ ms, Packet Loss Rate: $10^{-5}$ (TCP), $10^{-4}$ (UDP), Resolution: 3K, Coverage Level CSI RSRP: $-107$ dBm, Coverage Quality CSI SINR: 2 dB.\\
\hline
LLM (no-RAG) Output & Scenario Type: 4K On Demand Video
Key Performance Factors:
1. Video Quality: The video quality should be at least 3840x2160 pixels to ensure a true 4K experience.
2. Buffering Time: The buffering time should be minimal to provide a seamless viewing experience.
3. Playback Speed: The video should play at a consistent speed without any lag or delays.
4. Audio Quality: The audio quality should be clear and synchronized with the video.
5. User Interface: The user interface should be user-friendly and easy to navigate.
6. Content Selection: The platform should offer a wide variety of 4K content to attract and retain viewers.
7. Device Compatibility: The platform should be compatible with different devices to reach a larger audience.
8. Customer Support: The platform should have a reliable customer support system in place to address any technical issues.
9. Subscription Options: The platform should offer flexible subscription options to cater to different user preferences.
10. Data Usage: The platform should optimize data usage to prevent excessive data consumption for viewers. & Scenario Type: Virtual Reality Gaming

Key Performance Factors/Performance Recommendations:
1. Frame Rate: The frame rate of the game should be at least 90 frames per second to ensure smooth and realistic movement in the virtual environment. 
2. Latency: The latency, or delay, between the player's actions and the game's response should be minimal to avoid disorientation and motion sickness. 
3. Field of View: The field of view should be set to a comfortable level, typically between 90-110 degrees, to avoid a feeling of tunnel vision and motion sickness. 
4. Motion Blur: Motion blur should be minimized or turned off completely to reduce the strain on the player's eyes and prevent motion sickness. 
5. Head Tracking: The game should have accurate head tracking to ensure that the virtual environment moves in sync with the player's head movements, reducing the risk of motion sickness. 
6. Comfort Settings: The game should have options for comfort settings, such as reducing camera movement or adding a virtual nose, to help alleviate motion sickness for players who are more sensitive. 
7. Graphics Quality: The graphics quality should be optimized to ensure a smooth and realistic experience without causing lag or stuttering, which can contribute to motion sickness. 
8. Audio: The game's audio should be synchronized with the visuals to avoid any discrepancies that can cause disorientation and motion sickness. 
9. Breaks: The game should have built-in breaks or prompts for players to take breaks and rest their eyes to prevent eye strain and motion sickness. 
10. User Feedback: The game should have a system for users to provide feedback on their experience, allowing developers to make necessary adjustments to reduce the risk of motion sickness.\\
\hline
Ground Truth & Scenario Type: 4K On Demand Video
Key Performance Factors: Data Rate/Throughput (downlink): $30$ Mbps, Delay: RTT $< 100$ ms, Packet Loss Rate: $10^{-3}$, Resolution: $4$K, Coverage Level CSI RSRP:$-113$ dBm, Coverage Quality CSI SINR:$-2$ dB. & Scenario Type : 3K Cloud VR (Game),  
Key Performance Factors: Data Rate/Throughput (downlink): $100$ Mbps, Delay:RTT $< 25$ ms, Packet Loss Rate:$10^{-3}$ (TCP) $10^{-2}$ (UDP), Resolution: $3K$, Coverage Level CSI RSRP:$-107$ dBm, Coverage Quality CSI SINR:$2$ dB \\
\hline
\end{tabular}
\label{table_2}
\end{table*}
\begin{multicols}{2}
\end{multicols}

\begin{figure}
\centering
\includegraphics[width =3.5 in, height = 2.5in]{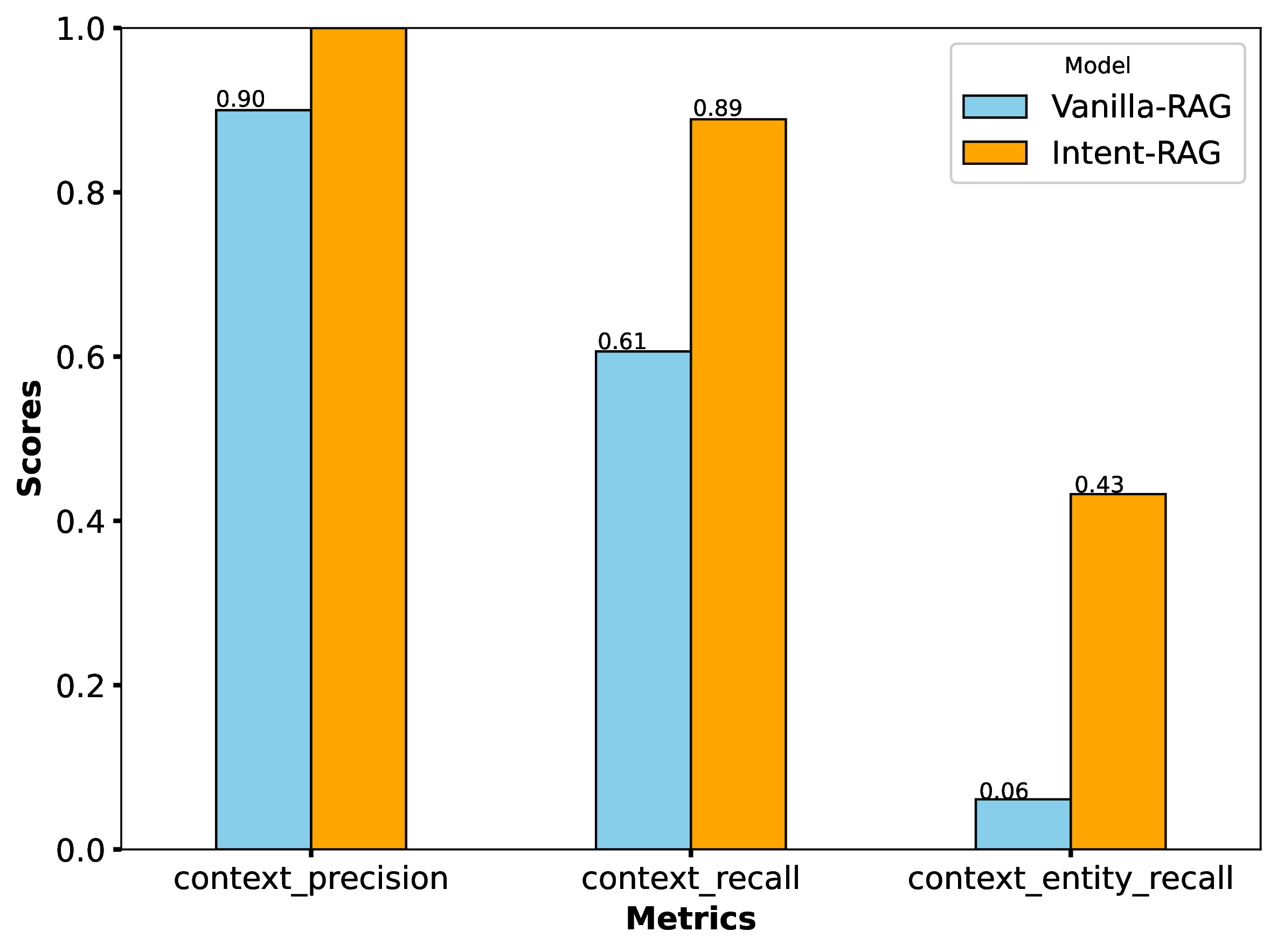}
\caption{Retrieval part performance of vanilla-RAG and intent-RAG.}
\label{Fig3}
\end{figure}

\begin{figure}
\centering
\includegraphics[width = 3.5 in, height = 2.5in]{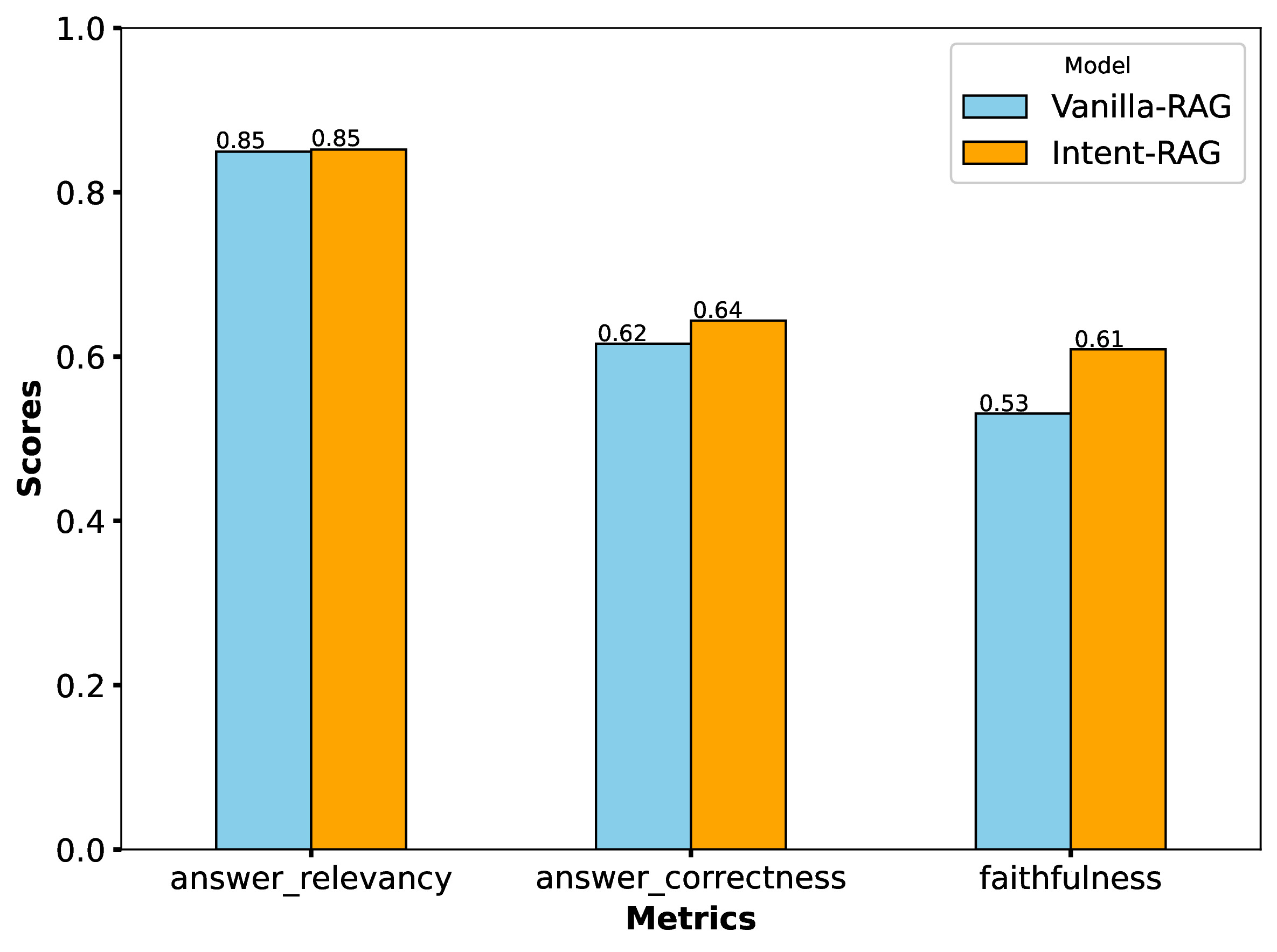}
\caption{Generative part performance of vanilla-RAG and intent-RAG.}
\label{Fig4}
\end{figure}

\begin{table}[h]
    \centering
    \caption{Translation time (second)} 
    \begin{tabular}{|c|c|c|} 
        \hline
         Vanilla-RAG & Intent-RAG & LLM (no-RAG)  \\  
        \hline
        2.16 & 2.62 & 3.33\\ 
        \hline
    \end{tabular}
    \label{table_3}
\end{table}

Figure~\ref{Fig3} demonstrates a comparison between vanilla-RAG and intent-RAG performance, focusing on the retrieval part using the RAGAS evaluator. As we can observe, the intent RAG outperforms vanilla-RAG in retrieving the relevant context for an expressed intent. The reason is that the intent RAG efficiently processes the network documents using unstructured~\cite{unstructured} compared to vanilla RAG. Moreover, the re-ranking model re-evaluates the retrieved context and reorders them based on relevancy compared to vanilla RAG. Figure~\ref{Fig4} illustrates a comparison between the performance of vanilla-RAG and intent-RAG, focusing on the generative part using the RAGAS evaluator. As we can see, both frameworks give close performance since the same LLM model is used in the generative part. We also compare it to an LLM (no-RAG) model, where a prompt and LLM are only used to translate the intent. We find that in no-RAG model the answer correctness was 0.29 only giving very poor performance. Table~\ref{table_2} shows the response from different frameworks. As we can see, intent-RAG outperforms vanilla-RAG in matching the ground truth while LLM (no-RAG) gives irrelevant answers. Table~\ref{table_3} demonstrates the average translation time over the data set. As we can observe, vanilla RAG has the shortest time in translation at the tradeoff of accuracy, while intent-RAG has a slightly longer translation time but with high accuracy. The slightly longer time in intent-RAG compared to vanilla RAG is due to the more processing required in intent-RAG framework stages. On the other hand, LLM (no-RAG) has the longest time due to the time required to search the public database.

\section{Conclusion}\label{conclusion}

In this paper, we investigated the application network interaction problem, where network user applications express their intents to the network in their domain languages (i.e., technical and non-technical) and the network interprets and translates them into structured network intents. We proposed a context-aware AI framework that utilizes machine reasoning, retrieval augmented generation, and generative AI technologies to build an intent translator. The proposed framework allows an understanding of the requirements of different types of applications (e.g., VR, AR, etc.) compared to many existing intent works, which focus on technical expressed intents. Experimental results show the intent RAG framework outperforms LLM and the vanilla RAG benchmark. In future work, we plan to utilize fine-tuned LLM models for network management. 

%{\color{blue} In summary, the formal problem is to learn the optimal mapping \( f_\theta \) that translates any user/application-level intent \( i \in \mathcal{I} \) into a structured, schema-compliant network intent \( n \in \mathcal{N} \), while minimizing translation error and maintaining high retrieval confidence.}
\section*{ACKNOWLEDGMENT}
 
The work is funded by the European Union through the project 6G-INTENSE (G.A no. 101139266). Views and opinions expressed are, however, those of the author(s) only and do not necessarily reflect those of the European Union. Neither the European Union nor the granting authority can be held responsible for them. 

\bibliographystyle{ieeetr}
\bibliography{References}

\end{document}